\documentstyle[12pt,epsfig,rotating,axodraw,psfrag,here]{article}

\catcode`@=11
\def\citer{\@ifnextchar
[{\@tempswatrue\@citexr}{\@tempswafalse\@citexr[]}}

\def\@citexr[#1]#2{\if@filesw\immediate\write\@auxout{\string\citation{#2}}\fi
  \def\@citea{}\@cite{\@for\@citeb:=#2\do
    {\@citea\def\@citea{--\penalty\@m}\@ifundefined
       {b@\@citeb}{{\bf ?}\@warning
       {Citation `\@citeb' on page \thepage \space undefined}}%
\hbox{\csname b@\@citeb\endcsname}}}{#1}}
\catcode`@=12

\newcommand{\MS}{\overline{\rm MS}}

\newcommand{\beq}{\begin{equation}}
\newcommand{\eeq}{\end{equation}}
\newcommand{\beqn}{\begin{eqnarray}}
\newcommand{\eeqn}{\end{eqnarray}}

\begin{document}

\renewcommand{\thefootnote}{\fnsymbol{footnote}}

\begin{flushright}
hep--ex/0302040 \\[1.0cm]
\end{flushright}

\begin{center}
{\Large \bf Photoproduction of $W$ Bosons at HERA: \\[0.5cm]
            Reweighting Method for implementing QCD Corrections
            in Monte Carlo Programs
} \\[1.0cm]

{\sc Kai--Peer O.~Diener$^1$, Christian
  Schwanenberger$^{2}$ and Michael Spira$^1$}\\[1.0cm]
{\it $^1$ Paul Scherrer Institut PSI, CH--5232 Villigen PSI, Switzerland\\
 $^2$ Deutsches Elektronen-Synchrotron DESY, D--22603 Hamburg,
 Germany} \\[1.0cm]
\end{center}

\begin{abstract}
\noindent
A procedure of implementing QCD corrections in Monte Carlo programs by a
reweighting method is described for the photoproduction of $W$ bosons at
HERA. Tables for $W$ boson production in LO and NLO are given in bins of
the transverse momentum of the W boson and its rapidity.
\end{abstract}

\renewcommand{\thefootnote}{\arabic{footnote}}
\setcounter{footnote}{0}

\section{Introduction}

In this note we describe the reweighting of leading order (LO) QCD Monte
Carlo (MC) programs for $W$ production using analytical next-to-leading
order (NLO) calculations~\citer{NRS,DSS} of the leading QCD corrections.
Since collinear divergences appear even at LO due to photon splitting
into collinear $q\bar q$ pairs, the reweighting has to be performed
differently for large and small values of $W$-transverse momenta
$p_{t,W}$.

First, a short overview over the theoretical status is given in Section
2.  The reweighting method is explained in Section 3 with particular
emphasis on the LO MC program EPVEC~\cite{baur}. In the appendices
tables for the reweighting are presented.

\section{NLO QCD Corrections to $W$ Production at HERA}

\subsection{Cross Sections for $W$ Bosons with transverse momentum}

\begin{figure}[h]
  \begin{center}
    \epsfig{file=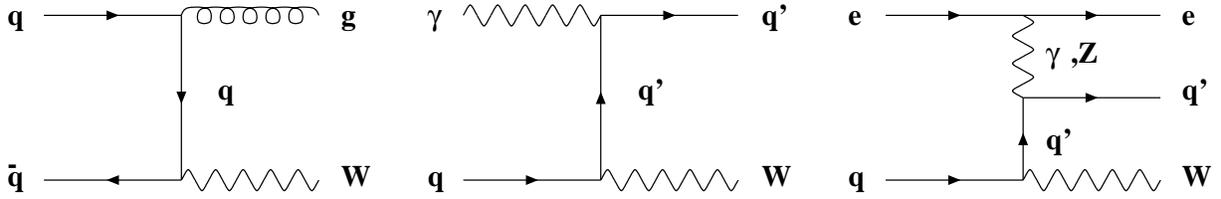,width=\hsize,bbllx=0,bblly=70,bburx=438,bbury=0}\vspace{3.3cm}
    \caption{\it
Typical LO diagrams of $W$ boson production with finite transverse
    momentum: resolved, direct and DIS mechanism.
    \label{fig:w1jet_LOgraphs}}
  \end{center}
\end{figure}

\begin{figure}[h]
  \begin{center}
    \epsfig{file=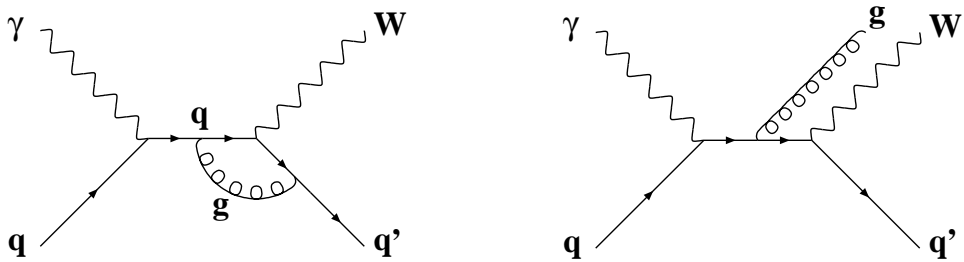,width=0.8\hsize,bbllx=0,bblly=86,bburx=312,bbury=0}\vspace{4.3cm}
    \caption{\it
Typical NLO diagrams (virtual and real corrections) of direct
    photoproduction of $W$ bosons with finite transverse 
    momentum.
    \label{fig:dir_NLOgraphs}}
  \end{center}
\end{figure}

In \cite{DSS} the differential cross section for 
\mbox{$e^\pm + p \rightarrow e^\pm +  W+X$}
(\mbox{$X= 1\,{\rm or}\,2\,{\rm jets}$}) has been calculated with respect to
the transverse momentum $p_{t,W}$ and rapidity $y_W$ of the $W$ boson;
resolved photoproduction is calculated in LO, direct photoproduction in
NLO QCD and deep inelastic scattering (DIS) in LO. Typical LO diagrams
for the three \mbox{$W+1\,{\rm jet}$} production mechanisms are depicted in
Fig.~\ref{fig:w1jet_LOgraphs}. Typical NLO diagrams in direct
photoproduction are shown in Fig.~\ref{fig:dir_NLOgraphs}.

For $p_{t,W} > 10$~GeV direct photoproduction is the dominating process.
NLO QCD corrections modify the LO direct cross section by about $\pm
(10-15)\%$ at the nominal renormalization/factorization scale
$\mu_R=\mu_F=M_W$ with $M_W$ being the $W$ mass \cite{DSS}. The
remaining theoretical uncertainty is estimated to be about 10\% (instead
of about 30\% in LO).

Since the NLO corrections are moderate and only hardly affect the shapes
of the differential distributions~\cite{DSS}, they can be implemented
with sufficient accuracy in a LO Monte Carlo (MC) program by reweighting
the generated events.  The hadronic parts of the processes are treated
inclusively so that no double counting arises with parton shower
effects. The procedure will be described in Section~\ref{sec:method}. 

\subsection{Total Cross Sections}

\begin{figure}[h]
  \begin{center}
    \epsfig{file=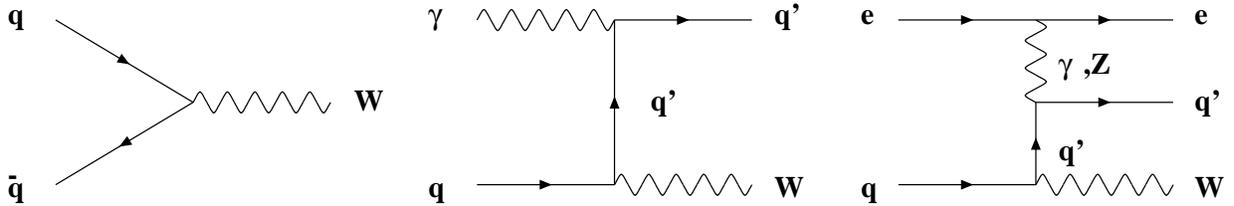,width=\hsize,bbllx=0,bblly=70,bburx=438,bbury=0}\vspace{3.3cm}
\caption{\it
Typical LO diagrams contributing to the total $W$ boson production cross
section: resolved, direct and DIS mechanism.
    \label{fig:wtot_LOgraphs}}
  \end{center}
\end{figure}

\begin{figure}[h]
  \begin{center}
    \epsfig{file=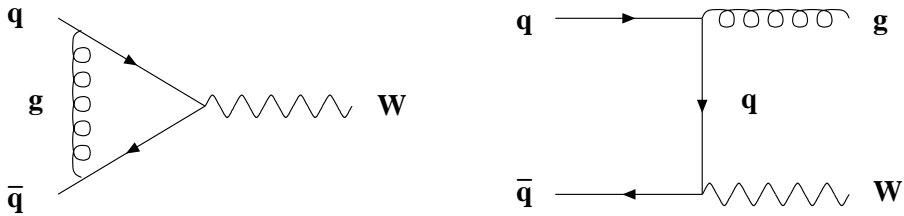,width=0.75\hsize,bbllx=0,bblly=85,bburx=358,bbury=0}\vspace{3.3cm}
    \caption{\it
Typical NLO diagrams (virtual and real corrections) contributing to the
    total $W$ boson production cross 
section in the resolved mechanism.
    \label{fig:res_NLOgraphs}}
  \end{center}
\end{figure}

In \cite{NRS,spira} the total cross section for resolved photoproduction
of $W$ bosons at HERA is calculated in NLO\footnote{Note that the LO
resolved photoproduction process  is of the order $\alpha_s^0$ for total
$W$ production (cf. Fig.~\ref{fig:wtot_LOgraphs}) while it is of first
order in $\alpha_s$ for \mbox{$W+1\,{\rm jet}$} production (cf.
Fig~\ref{fig:w1jet_LOgraphs}).}.  The total cross sections for direct
photoproduction and DIS are calculated in LO.  Typical LO diagrams for
these three types of $W$ boson production mechanisms are depicted in
Fig.~\ref{fig:wtot_LOgraphs}.  Typical NLO diagrams in resolved
photoproduction are shown in Fig.~\ref{fig:res_NLOgraphs}.

The total cross section is dominated by low $p_{t,W}$ contributions
coming from resolved photoproduction. For this process, the NLO QCD
corrections modify the LO contribution by about 40\% at the nominal
renormalization/factorization scale $\mu_R=\mu_F=M_W$, thus affecting
the total $W$ production rate significantly.
The inclusion of NLO QCD contribution is estimated to reduce the 
theoretical uncertainty to roughly $10$\% compared to about $30$\% 
at LO.

\section{Reweighting Method}
\label{sec:method}

In \citer{NRS,DSS} the DIS and photoproduction regimes are separated by
a conventional cut in the photon virtuality $-Q^2$. The photoproduction
regime is defined as $Q^2 < Q^2_{\rm max}$ ($Q^2_{\rm max}$ chosen as
4~${\rm GeV}^2$). If a LO MC uses the same separation cut, the
reweighting can be performed separately for the photoproduction regime
(DIS stays unreweighted in that case), which is described in the
following.  This is, however, not the case for the EPVEC MC which will
be discussed separately.

\subsection{Reweighting for finite transverse momenta of the $W$ boson}

For finite transverse momenta of the $W$ boson the event sample for
generated partons in the LO MC is divided in bins of the $W$'s
transverse momentum $p_{t,W}$ and rapidity $y_W$ (defined to be positive
in the electron/positron direction).  To implement NLO corrections each
generated photoproduction event of the corresponding bin acquires a new
weight 
\begin{equation}
w = \frac{\displaystyle \frac{d^2\sigma_{res}^{LO}}{d p_{t,W} \, d y_W} +
\frac{d^2\sigma_{dir}^{NLO}}{d p_{t,W} \, d y_W}}{\displaystyle
\frac{d^2\sigma_{res}^{MC}}{d p_{t,W} \, d y_W} +
\frac{d^2\sigma_{dir}^{MC}}{d p_{t,W} \, d y_W}} \,\, ,
\label{weight}
\end{equation}
where $\frac{d^2\sigma_{res}^{LO}}{d p_{t,W} \, d y_W}$ and
$\frac{d^2\sigma_{dir}^{NLO}}{d p_{t,W} \, d y_W}$ are the double
differential cross sections for resolved photoproduction in LO and
direct photoproduction in NLO as calculated in~\cite{DSS}.
$\frac{d^2\sigma_{res}^{MC}}{d p_{t,W} \, d y_W}$ and
$\frac{d^2\sigma_{dir}^{MC}}{d p_{t,W} \, d y_W}$ are the double
differential cross sections for resolved and direct photoproduction in
LO taken from the MC\footnote{Note that the denominator in
Eq.~(\ref{weight}) involves the LO cross sections of the MC and {\em
not} of the analytical calculation~\cite{DSS}. This has the advantage
that the major part of differences between the analytical calculation in
LO and the MC which could emerge due to different conventions is
corrected for.}.

In the appendix tables for the double differential cross sections are
given for several values of $p_{t,W}$ and $y_W$ as calculated
in~\cite{DSS}. These values should be taken as mean values for the
respective bins. The bin widths are $\Delta p_{t,W} = 10$~GeV and
$\Delta y_W = 0.25$. The range of $5 < p_{t,W} < 105$~GeV and $-1.375 <
y_{W,cms} < 1.375$ in the center of mass system is covered. Bin
fluctuations are estimated be less than 2\% and thus negligible. 

\subsection{Reweighting for small transverse momenta of the $W$ boson}

Towards small transverse momenta $p_{t,W}$ the differential cross
sections diverge due to collinear and infrared divergences. Thus, only
the renormalized {\em total} cross sections are calculated
analytically~\cite{NRS,spira}. In the NLO calculation the
rapidity-integrated cross section for $p_{t,W} < 5$~GeV can be obtained
as
\begin{equation}
\sigma_{res}^{NLO}(p_{t,W} < 5\,{\rm GeV}) = \sigma_{res}^{NLO}({\rm total})
- \sigma_{res}^{LO}(p_{t,W} > 5\,{\rm GeV}) \label{reslt5gev}
\end{equation}
for resolved photoproduction\footnote{Note that both
$\sigma_{res}^{NLO}({\rm total})$ and $\sigma_{res}^{LO}(p_{t,W} >
5\,{\rm GeV})$ are of the same order of $\alpha_s$ (cf.
Figs.~\ref{fig:w1jet_LOgraphs},~\ref{fig:res_NLOgraphs}).} and by
\begin{equation}
\sigma_{dir}^{LO}(p_{t,W} < 5\,{\rm GeV}) = \sigma_{dir}^{LO}({\rm total}) -
\sigma_{dir}^{LO}(p_{t,W} > 5\,{\rm GeV})
\label{dirlt5gev}
\end{equation}
for direct photoproduction. The numerical values are given in the
appendix. However, it should be stressed that the values for
$\sigma_{res}^{LO}(p_{t,W} > 5\,{\rm GeV})$ and
$\sigma_{dir}^{LO}(p_{t,W}
> 5\,{\rm GeV})$ develop sizeable theoretical uncertainties. 
For the lowest bins with $p_{t,W} < 5$~GeV a rapidity-independent
weighting factor is applied:
\begin{eqnarray}
w(p_{t,W} < 5\,{\rm GeV}) &=& \frac{\sigma_{res}^{NLO}(p_{t,W} < 5\,{\rm
GeV}) + \sigma_{dir}^{LO}(p_{t,W} < 5\,{\rm GeV})}{\sigma_{res}^{MC}(p_{t,W}
< 5\,{\rm GeV}) + \sigma_{dir}^{MC}(p_{t,W} < 5\,{\rm GeV})} \,\, .
\label{weightlbin}
\end{eqnarray}

\subsection{Reweighting EPVEC MC}

The analytical calculations~\citer{NRS,DSS} differ from the calculations
of the EPVEC MC generator~\cite{baur} in several aspects which are
discussed in detail in~\cite{spira}. The differences are well understood
and of the order of 10\% in the total $W$ production cross section.

\begin{figure}[t]
   \begin{picture}(0,270)
     \psfrag{b}[][][1.4][0]{$\mu_F$}
     \psfrag{c}[][][1.4][0]{$Q^2_{max}$}
     \psfrag{d}[][][1.7][0]{$Q^2$}
     \psfrag{e}[][][1.7][0]{$|\hat{u}|$}
     \psfrag{f}[][][1.4][0]{$u_{cut}$}
     \put(-20,270){\epsfig{file=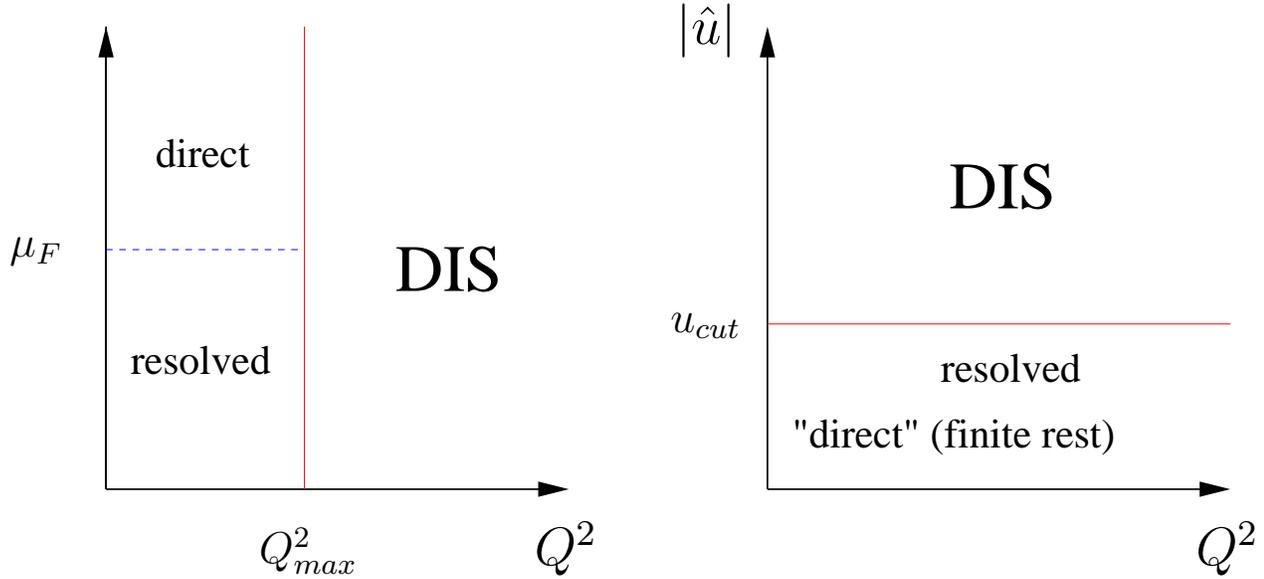,width=\hsize,bbllx=0,bblly=290,bburx=655,bbury=0}}
   \end{picture}
\caption[]{\it
Different separation between DIS and photoproduction regime in the
analytical calculations of Refs.~\citer{NRS,DSS} (left) and
EPVEC~\cite{baur} (right). \label{fig:sep}}
\vspace{0.2cm}
\end{figure}

One main difference between EPVEC and the above analytical calculations
is the different separation between DIS and photoproduction regimes.
EPVEC imposes a cut on the $u$-channel momentum transfer in the
$\gamma^*q$ subprocess (see second diagram of
Figs.~\ref{fig:w1jet_LOgraphs},~\ref{fig:wtot_LOgraphs}) for the
separation. In \citer{NRS,DSS} the separation is achieved by means of a
cut in the photon virtuality $-Q^2$. These differences are qualitatively
illustrated in Fig.~\ref{fig:sep}\footnote{In Refs.~\citer{NRS,DSS}
direct and resolved photoproduction are disentangled by mass
factorization in dimensional regularization. This means, that collinear
singularities, arising for the final state quark $q'$ being collinear
with the initial state photon, are absorbed into the photonic quark
densities defined at the factorization scale $\mu_F$.

In EPVEC the resulting $\gamma q \rightarrow W q'$ cross section for the
region $|\hat{u}| < u_{cut}$ is split into three parts (cf. Eq.~(2.9) of
\cite{baur}): one contains the photon structure function and is
therefore equivalent to the resolved photon part.  The residual two
parts determine the finite rest after factorizing the resolved photon
contribution and remove double counting of the resolved photon cross 
section with the DIS region. 
Both terms involve direct photon couplings to fermions and
$W$ bosons.  However, this should {\em not} be mixed up with direct
photoproduction as defined in Refs.~\citer{NRS,DSS} and in usual HERA
analyses.  Moreover, in EPVEC the resolved photon part is defined in the
DIS$_\gamma$ scheme as opposed to the $\overline{\rm MS}$ scheme of the
resolved photoproduction part in Refs.~\citer{NRS,DSS} and in usual HERA
analyses. 

The deep inelastic region in EPVEC, defined by $|\hat{u}| > u_{cut}$, is
also {\em not} identical to the DIS part in Ref.~\cite{DSS} and usual
HERA analyses. In particular, the direct photoproduction events of
Refs.~\citer{NRS,DSS} with $|\hat{u}| > u_{cut}$ are included in the
deep inelastic part of EPVEC.}.

As a consequence only the sum of DIS and photoproduction cross sections
can be reweighted. The weighting factor corresponding to
Eq.~(\ref{weight}) which has now to be applied to each generated EPVEC W
production event (DIS {\em and} photoproduction), reads
\begin{eqnarray}
w = \frac{\displaystyle \frac{d^2\sigma_{res}^{LO}}{d p_{t,W} \, d y_W} +
\frac{d^2\sigma_{dir}^{NLO}}{d p_{t,W} \, d y_W} +
\frac{d^2\sigma_{DIS}^{LO}}{d p_{t,W} \, d y_W}} {\displaystyle
\frac{d^2\sigma_{W\, production}^{\mathrm EPVEC}}{d p_{t,W} \, d y_W}} \,\, ,
\label{weightEPVEC}
\end{eqnarray}
with the double differential cross section for DIS in LO
$\frac{d^2\sigma_{DIS}^{LO}}{d p_{t,W} \, d y_W}$ as calculated
in~\cite{DSS} and the W production cross section $\frac{d^2\sigma_{W\,
production}^{{\mathrm EPVEC}}}{d p_{t,W} \, d y_W}$ taken from
EPVEC\footnote{Corresponding to the previous footnote, the W production
cross section in EPVEC is the sum of the deep inelastic region with
$|\hat{u}| > u_{cut}$ and the photoproduction region with $|\hat{u}| <
u_{cut}$.}.  Using the EPVEC LO cross sections in the
denominator of the weighting factors corrects for the major differences
between EPVEC conventions and those of Ref.~\cite{DSS}. (The use of the
LO cross sections of \cite{DSS} would be inconsistent.)

The weighting factor for the first $p_{t,W}$ bin corresponding to
Eq.~(\ref{weightlbin}) is given by
\begin{equation}
w(p_{t,W} \! < \! 5\,{\rm GeV}) = \frac{\displaystyle
\sigma_{res}^{NLO}(p_{t,W} \! < \! 5\,{\rm GeV}) +
\sigma_{dir}^{LO}(p_{t,W} \! < \!  5\,{\rm GeV})
+\sigma_{DIS}^{LO}(p_{t,W} \! < \! 5\,{\rm GeV})} {\displaystyle
\sigma_{W\, production}^{\big. {\mathrm EPVEC} \big. }(p_{t,W} \! < \! 5\,{\rm GeV})} \, .
\label{weightlbinEPVEC} \\
\end{equation}

\noindent
{\bf Acknowledgements} \\
We are grateful to D.\,Dannheim, C.\,Diaconu, and M.\,Schneider for
useful discussions.


\newpage
\begin{appendix}

\section{Double Differential Cross Sections}

Double differential cross sections for different values of transverse
momenta $p_{t,W}$ and rapidity, both in the laboratory ($y_{W,lab}$) and
the center of mass frame ($y_{W,cms}$), are presented for the individual
$W$ production processes calculated in LO and NLO, respectively. The
results are given for proton beam energies of $E_p=820$~GeV and
$E_p=920$~GeV (and electron/positron beam energies of $E_e = 27.5$~GeV).
The ACFGP~\cite{acfgp} parton densities are chosen for the photon.   All
other settings and numerical values of parameters and are as
in~\cite{DSS}. No entry means that the respective point is out of phase
space.

\subsection{Double Differential Cross Sections in LO}

For the proton CTEQ4L densities~\cite{cteq} are used with LO strong
coupling ($\Lambda^{(LO)}_5=181$~MeV).  The following values can be used
for consistency checks in LO.

\subsubsection{$W^+$ production in $e^+p$ scattering in LO at $E_p=920$~GeV}

\begin{table}[H]
  \begin{center}
\begin{tabular}{|c|c|c||c|c|c||c|}
\hline
$p_{t,W}$ (GeV) & $y_{W,lab}$  & $y_{W,cms}$ & resolved & direct & DIS &
total \\
\hline \hline
 10. & -3.005 & -1.25 & 0.8131E-06 & 0.1855E-05 & 0.5255E-06 & 0.3194E-05 \\
 10. & -2.755 & -1.00 & 0.1043E-03 & 0.1704E-03 & 0.4724E-04 & 0.3219E-03 \\ 
 10. & -2.505 & -0.75 & 0.5695E-03 & 0.8862E-03 & 0.2477E-03 & 0.1703E-02 \\
 10. & -2.255 & -0.50 & 0.1324E-02 & 0.2107E-02 & 0.6025E-03 & 0.4033E-02 \\
 10. & -2.005 & -0.25 & 0.2089E-02 & 0.3507E-02 & 0.1033E-02 & 0.6629E-02 \\
 10. & -1.755 &  0.00 & 0.2645E-02 & 0.4772E-02 & 0.1452E-02 & 0.8869E-02 \\
 10. & -1.505 &  0.25 & 0.2881E-02 & 0.5696E-02 & 0.1811E-02 & 0.1039E-01 \\
 10. & -1.255 &  0.50 & 0.2759E-02 & 0.6170E-02 & 0.2055E-02 & 0.1098E-01 \\
 10. & -1.005 &  0.75 & 0.2268E-02 & 0.6142E-02 & 0.2192E-02 & 0.1060E-01 \\
 10. & -0.755 &  1.00 & 0.1396E-02 & 0.5494E-02 & 0.2175E-02 & 0.9065E-02 \\
 10. & -0.505 &  1.25 & 0.2684E-03 & 0.3588E-02 & 0.1814E-02 & 0.5670E-02 \\
\hline                                                       
 20. & -3.005 & -1.25 & 0.2323E-07 & 0.1391E-06 & 0.5303E-07 & 0.2154E-06 \\
 20. & -2.755 & -1.00 & 0.1120E-04 & 0.4182E-04 & 0.1516E-04 & 0.6818E-04 \\
 20. & -2.505 & -0.75 & 0.8395E-04 & 0.3049E-03 & 0.1093E-03 & 0.4981E-03 \\
 20. & -2.255 & -0.50 & 0.2262E-03 & 0.8869E-03 & 0.3219E-03 & 0.1435E-02 \\
 20. & -2.005 & -0.25 & 0.3876E-03 & 0.1711E-02 & 0.6338E-03 & 0.2732E-02 \\
 20. & -1.755 &  0.00 & 0.5144E-03 & 0.2620E-02 & 0.1002E-02 & 0.4136E-02 \\
 20. & -1.505 &  0.25 & 0.5728E-03 & 0.3456E-02 & 0.1377E-02 & 0.5406E-02 \\
 20. & -1.255 &  0.50 & 0.5464E-03 & 0.4090E-02 & 0.1707E-02 & 0.6343E-02 \\
 20. & -1.005 &  0.75 & 0.4283E-03 & 0.4420E-02 & 0.1965E-02 & 0.6813E-02 \\
 20. & -0.755 &  1.00 & 0.2246E-03 & 0.4291E-02 & 0.2087E-02 & 0.6603E-02 \\
 20. & -0.505 &  1.25 & 0.1828E-04 & 0.2805E-02 & 0.1632E-02 & 0.4455E-02 \\
\hline                            
\end{tabular}
  \end{center}
\end{table}

\newpage
\thispagestyle{empty}
\begin{table}[H]
  \begin{center}
\begin{tabular}{|c|c|c||c|c|c||c|}
\hline
$p_{t,W}$ (GeV) & $y_{W,lab}$  & $y_{W,cms}$ & resolved & direct & DIS &
total \\
\hline \hline
 30. & -3.005 & -1.25 & 0.2282E-09 & 0.3310E-08 & 0.1568E-08 & 0.5106E-08 \\
 30. & -2.755 & -1.00 & 0.1531E-05 & 0.1068E-04 & 0.4488E-05 & 0.1670E-04 \\
 30. & -2.505 & -0.75 & 0.1736E-04 & 0.1176E-03 & 0.4860E-04 & 0.1836E-03 \\
 30. & -2.255 & -0.50 & 0.5570E-04 & 0.4223E-03 & 0.1755E-03 & 0.6535E-03 \\
 30. & -2.005 & -0.25 & 0.1046E-03 & 0.9393E-03 & 0.3940E-03 & 0.1438E-02 \\
 30. & -1.755 &  0.00 & 0.1459E-03 & 0.1602E-02 & 0.6883E-03 & 0.2436E-02 \\
 30. & -1.505 &  0.25 & 0.1657E-03 & 0.2300E-02 & 0.1018E-02 & 0.3484E-02 \\
\hline                            
\end{tabular}
  \end{center}
\end{table}

\newpage
\thispagestyle{empty}
\begin{table}[H]
  \begin{center}
\begin{tabular}{|c|c|c||c|c|c||c|}
\hline
$p_{t,W}$ (GeV) & $y_{W,lab}$  & $y_{W,cms}$ & resolved & direct & DIS &
total \\
\hline \hline
 70. & -3.005 & -1.25 &   ---      &   ---      &   ---      &   ---      \\
 70. & -2.755 & -1.00 & 0.3952E-12 & 0.5213E-10 & 0.3583E-10 & 0.8836E-10 \\
 70. & -2.505 & -0.75 & 0.2090E-07 & 0.8048E-06 & 0.4760E-06 & 0.1302E-05 \\
 70. & -2.255 & -0.50 & 0.3068E-06 & 0.1289E-04 & 0.7302E-05 & 0.2050E-04 \\
 70. & -2.005 & -0.25 & 0.1065E-05 & 0.5844E-04 & 0.3245E-04 & 0.9196E-04 \\
 70. & -1.755 &  0.00 & 0.1935E-05 & 0.1530E-03 & 0.8445E-04 & 0.2394E-03 \\
 70. & -1.505 &  0.25 & 0.2279E-05 & 0.2890E-03 & 0.1587E-03 & 0.4500E-03 \\
\hline                            
\hline                            
\end{tabular}
\parbox{0.9\textwidth}{\caption{\it
{${d^2\sigma}/{d p_{t,W} \, d y_W}$ (in $pb$/GeV) in LO.}}}
  \end{center}
\end{table}

\subsubsection{$W^-$ production in $e^+p$ scattering in LO at $E_p=920$~GeV}

\begin{table}[H]
  \begin{center}
\begin{tabular}{|c|c|c||c|c|c||c|}
\hline
$p_{t,W}$ (GeV) & $y_{W,lab}$  & $y_{W,cms}$ & resolved & direct & DIS &
total \\
\hline \hline
 10. & -3.005 & -1.25 & 0.1793E-06 & 0.4959E-06 & 0.1413E-06 & 0.8165E-06 \\
 10. & -2.755 & -1.00 & 0.4464E-04 & 0.8154E-04 & 0.2281E-04 & 0.1490E-03 \\
 10. & -2.505 & -0.75 & 0.3561E-03 & 0.5749E-03 & 0.1620E-03 & 0.1093E-02 \\
 10. & -2.255 & -0.50 & 0.1074E-02 & 0.1661E-02 & 0.4746E-03 & 0.3210E-02 \\
 10. & -2.005 & -0.25 & 0.2034E-02 & 0.3139E-02 & 0.9106E-03 & 0.6084E-02 \\
 10. & -1.755 &  0.00 & 0.2914E-02 & 0.4619E-02 & 0.1360E-02 & 0.8893E-02 \\
 10. & -1.505 &  0.25 & 0.3429E-02 & 0.5750E-02 & 0.1723E-02 & 0.1090E-01 \\
\hline                            
\end{tabular}
  \end{center}
\end{table}

\newpage
\thispagestyle{empty}
\begin{table}[H]
  \begin{center}
\begin{tabular}{|c|c|c||c|c|c||c|}
\hline
$p_{t,W}$ (GeV) & $y_{W,lab}$  & $y_{W,cms}$ & resolved & direct & DIS &
total \\
\hline \hline
 40. & -3.005 & -1.25 &   ---      &   ---      &   ---      &   ---      \\
 40. & -2.755 & -1.00 & 0.5961E-07 & 0.7095E-06 & 0.3342E-06 & 0.1103E-05 \\
 40. & -2.505 & -0.75 & 0.1896E-05 & 0.1879E-04 & 0.8473E-05 & 0.2916E-04 \\
 40. & -2.255 & -0.50 & 0.1011E-04 & 0.9964E-04 & 0.4413E-04 & 0.1539E-03 \\
 40. & -2.005 & -0.25 & 0.2604E-04 & 0.2783E-03 & 0.1223E-03 & 0.4266E-03 \\
 40. & -1.755 &  0.00 & 0.4447E-04 & 0.5497E-03 & 0.2414E-03 & 0.8356E-03 \\
 40. & -1.505 &  0.25 & 0.5670E-04 & 0.8694E-03 & 0.3810E-03 & 0.1307E-02 \\
\hline                            
\end{tabular}
  \end{center}
\end{table}

\begin{table}[H]
  \begin{center}
\begin{tabular}{|c|c|c||c|c|c||c|}
\hline
$p_{t,W}$ (GeV) & $y_{W,lab}$  & $y_{W,cms}$ & resolved & direct & DIS &
total \\
\hline \hline
 80. & -3.005 & -1.25 &   ---      &   ---      &   ---      &   ---      \\
 80. & -2.755 & -1.00 &   ---      &   ---      &   ---      &   ---      \\
 80. & -2.505 & -0.75 & 0.5040E-09 & 0.2480E-07 & 0.1443E-07 & 0.3973E-07 \\
 80. & -2.255 & -0.50 & 0.3074E-07 & 0.1291E-05 & 0.7001E-06 & 0.2022E-05 \\
 80. & -2.005 & -0.25 & 0.1867E-06 & 0.8849E-05 & 0.4655E-05 & 0.1369E-04 \\
 80. & -1.755 &  0.00 & 0.4491E-06 & 0.2787E-04 & 0.1433E-04 & 0.4265E-04 \\
 80. & -1.505 &  0.25 & 0.5837E-06 & 0.5572E-04 & 0.2834E-04 & 0.8464E-04 \\
\hline                            
\hline                            
\end{tabular}
\parbox{0.9\textwidth}{\caption{\it
{${d^2\sigma}/{d p_{t,W} \, d y_W}$ (in $pb$/GeV) in LO.}}}
  \end{center}
\end{table}

\subsubsection{$W^+$ production in $e^+p$ scattering in LO at $E_p=820$~GeV}

\begin{table}[H]
  \begin{center}
\begin{tabular}{|c|c|c||c|c|c||c|}
\hline
$p_{t,W}$ (GeV) & $y_{W,lab}$  & $y_{W,cms}$ & resolved & direct & DIS &
total \\
\hline \hline
 10. & -2.948 & -1.25 & 0.3168E-07 & 0.9891E-07 & 0.2882E-07 & 0.1594E-06 \\
 10. & -2.698 & -1.00 & 0.4986E-04 & 0.8805E-04 & 0.2460E-04 & 0.1625E-03 \\
 10. & -2.448 & -0.75 & 0.3891E-03 & 0.6344E-03 & 0.1784E-03 & 0.1202E-02 \\
 10. & -2.198 & -0.50 & 0.1030E-02 & 0.1699E-02 & 0.4881E-03 & 0.3217E-02 \\
 10. & -1.948 & -0.25 & 0.1727E-02 & 0.2991E-02 & 0.8845E-03 & 0.5602E-02 \\
 10. & -1.698 &  0.00 & 0.2252E-02 & 0.4192E-02 & 0.1280E-02 & 0.7724E-02 \\
 10. & -1.448 &  0.25 & 0.2479E-02 & 0.5078E-02 & 0.1618E-02 & 0.9175E-02 \\
\hline                            
\end{tabular}
  \end{center}
\end{table}

\newpage
\thispagestyle{empty}
\begin{table}[H]
  \begin{center}
\begin{tabular}{|c|c|c||c|c|c||c|}
\hline
$p_{t,W}$ (GeV) & $y_{W,lab}$  & $y_{W,cms}$ & resolved & direct & DIS &
total \\
\hline \hline
 40. & -2.948 & -1.25 &   ---      &   ---      &   ---      &   ---      \\
 40. & -2.698 & -1.00 & 0.4123E-07 & 0.5774E-06 & 0.2788E-06 & 0.8974E-06 \\
 40. & -2.448 & -0.75 & 0.2008E-05 & 0.2418E-04 & 0.1124E-04 & 0.3743E-04 \\
 40. & -2.198 & -0.50 & 0.1021E-04 & 0.1371E-03 & 0.6307E-04 & 0.2104E-03 \\
 40. & -1.948 & -0.25 & 0.2365E-04 & 0.3889E-03 & 0.1797E-03 & 0.5923E-03 \\
 40. & -1.698 &  0.00 & 0.3644E-04 & 0.7737E-03 & 0.3624E-03 & 0.1173E-02 \\
 40. & -1.448 &  0.25 & 0.4254E-04 & 0.1234E-02 & 0.5916E-03 & 0.1868E-02 \\
\hline                            
\end{tabular}
  \end{center}
\end{table}

\begin{table}[H]
  \begin{center}
\begin{tabular}{|c|c|c||c|c|c||c|}
\hline
$p_{t,W}$ (GeV) & $y_{W,lab}$  & $y_{W,cms}$ & resolved & direct & DIS &
total \\
\hline \hline
 80. & -2.948 & -1.25 &   ---      &   ---      &   ---      &   ---      \\
 80. & -2.698 & -1.00 &   ---      &   ---      &   ---      &   ---      \\
 80. & -2.448 & -0.75 & 0.1504E-09 & 0.1291E-07 & 0.8212E-08 & 0.2127E-07 \\
 80. & -2.198 & -0.50 & 0.2459E-07 & 0.1747E-05 & 0.1039E-05 & 0.2811E-05 \\
 80. & -1.948 & -0.25 & 0.1581E-06 & 0.1437E-04 & 0.8241E-05 & 0.2277E-04 \\
 80. & -1.698 &  0.00 & 0.3519E-06 & 0.4871E-04 & 0.2747E-04 & 0.7653E-04 \\
 80. & -1.448 &  0.25 & 0.4027E-06 & 0.1015E-03 & 0.5691E-04 & 0.1588E-03 \\
\hline                            
\hline                            
\end{tabular}
\parbox{0.9\textwidth}{\caption{\it
{${d^2\sigma}/{d p_{t,W} \, d y_W}$ (in $pb$/GeV) in LO.}}}
  \end{center}
\end{table}

\subsubsection{$W^-$ production in $e^+p$ scattering in LO at $E_p=820$~GeV}

\begin{table}[H]
  \begin{center}
\begin{tabular}{|c|c|c||c|c|c||c|}
\hline
$p_{t,W}$ (GeV) & $y_{W,lab}$  & $y_{W,cms}$ & resolved & direct & DIS &
total \\
\hline \hline
 10. & -2.948 & -1.25 & 0.5473E-08 & 0.2069E-07 & 0.6063E-08 & 0.3223E-07 \\
 10. & -2.698 & -1.00 & 0.1930E-04 & 0.3815E-04 & 0.1074E-04 & 0.6819E-04 \\
 10. & -2.448 & -0.75 & 0.2288E-03 & 0.3872E-03 & 0.1096E-03 & 0.7256E-03 \\
 10. & -2.198 & -0.50 & 0.8009E-03 & 0.1284E-02 & 0.3683E-03 & 0.2453E-02 \\
 10. & -1.948 & -0.25 & 0.1633E-02 & 0.2600E-02 & 0.7565E-03 & 0.4990E-02 \\
 10. & -1.698 &  0.00 & 0.2432E-02 & 0.3975E-02 & 0.1171E-02 & 0.7578E-02 \\
 10. & -1.448 &  0.25 & 0.2915E-02 & 0.5052E-02 & 0.1516E-02 & 0.9483E-02 \\
\hline                            
\end{tabular}
  \end{center}
\end{table}

\newpage
\thispagestyle{empty}
\begin{table}[H]
  \begin{center}
\begin{tabular}{|c|c|c||c|c|c||c|}
\hline
$p_{t,W}$ (GeV) & $y_{W,lab}$  & $y_{W,cms}$ & resolved & direct & DIS &
total \\
\hline \hline
 40. & -2.948 & -1.25 &   ---      &   ---      &   ---      &   ---      \\
 40. & -2.698 & -1.00 & 0.1162E-07 & 0.1633E-06 & 0.7834E-07 & 0.2533E-06 \\
 40. & -2.448 & -0.75 & 0.9161E-06 & 0.9828E-05 & 0.4477E-05 & 0.1522E-04 \\
 40. & -2.198 & -0.50 & 0.6357E-05 & 0.6625E-04 & 0.2956E-04 & 0.1022E-03 \\
 40. & -1.948 & -0.25 & 0.1838E-04 & 0.2063E-03 & 0.9101E-04 & 0.3157E-03 \\
 40. & -1.698 &  0.00 & 0.3315E-04 & 0.4319E-03 & 0.1897E-03 & 0.6547E-03 \\
 40. & -1.448 &  0.25 & 0.4300E-04 & 0.7050E-03 & 0.3102E-03 & 0.1058E-02 \\
\hline                            
\end{tabular}
  \end{center}
\end{table}

\begin{table}[H]
  \begin{center}
\begin{tabular}{|c|c|c||c|c|c||c|}
\hline
$p_{t,W}$ (GeV) & $y_{W,lab}$  & $y_{W,cms}$ & resolved & direct & DIS &
total \\
\hline \hline
 80. & -2.948 & -1.25 &   ---      &   ---      &   ---      &   ---      \\
 80. & -2.698 & -1.00 &   ---      &   ---      &   ---      &   ---      \\
 80. & -2.448 & -0.75 & 0.3712E-10 & 0.2587E-08 & 0.1531E-08 & 0.4155E-08 \\
 80. & -2.198 & -0.50 & 0.1021E-07 & 0.5057E-06 & 0.2752E-06 & 0.7911E-06 \\
 80. & -1.948 & -0.25 & 0.8823E-07 & 0.4802E-05 & 0.2507E-05 & 0.7397E-05 \\
 80. & -1.698 &  0.00 & 0.2378E-06 & 0.1715E-04 & 0.8814E-05 & 0.2620E-04 \\
 80. & -1.448 &  0.25 & 0.3046E-06 & 0.3535E-04 & 0.1798E-04 & 0.5363E-04 \\
\hline                            
\hline                            
\end{tabular}
\parbox{0.9\textwidth}{\caption{\it
{${d^2\sigma}/{d p_{t,W} \, d y_W}$ (in $pb$/GeV) in LO.}}}
  \end{center}
\end{table}

\newpage

\subsubsection{$W^\pm$ production in $e^-p$ scattering in LO at $E_p=920$~GeV}

In $e^-p$ scattering the results for resolved and direct photoproduction
are identical to those of $e^+p$ scattering, while the DIS results are
different:

\begin{table}[H]
  \begin{center}
\begin{tabular}{|c|c|c||c|c|}
\hline
$p_{t,W}$ (GeV) & $y_{W,lab}$  & $y_{W,cms}$  & DIS: $W^+$ production & DIS:
$W^-$ production \\
\hline \hline
 10. & -3.005 & -1.25 & 0.5293E-06 & 0.1421E-06 \\
 10. & -2.755 & -1.00 & 0.4723E-04 & 0.2297E-04 \\
 10. & -2.505 & -0.75 & 0.2459E-03 & 0.1629E-03 \\
 10. & -2.255 & -0.50 & 0.5924E-03 & 0.4762E-03 \\
 10. & -2.005 & -0.25 & 0.1003E-02 & 0.9137E-03 \\
 10. & -1.755 &  0.00 & 0.1400E-02 & 0.1371E-02 \\
 10. & -1.505 &  0.25 & 0.1713E-02 & 0.1738E-02 \\
 10. & -1.255 &  0.50 & 0.1904E-02 & 0.1950E-02 \\
 10. & -1.005 &  0.75 & 0.1963E-02 & 0.1999E-02 \\
 10. & -0.755 &  1.00 & 0.1863E-02 & 0.1857E-02 \\
 10. & -0.505 &  1.25 & 0.1423E-02 & 0.1301E-02 \\
 \hline
 20. & -3.005 & -1.25 & 0.5471E-07 & 0.1275E-07 \\
 20. & -2.755 & -1.00 & 0.1545E-04 & 0.6544E-05 \\
 20. & -2.505 & -0.75 & 0.1105E-03 & 0.6238E-04 \\
 20. & -2.255 & -0.50 & 0.3184E-03 & 0.2142E-03 \\
 20. & -2.005 & -0.25 & 0.6188E-03 & 0.4610E-03 \\
 20. & -1.755 &  0.00 & 0.9592E-03 & 0.7611E-03 \\
 20. & -1.505 &  0.25 & 0.1286E-02 & 0.1052E-02 \\
 20. & -1.255 &  0.50 & 0.1546E-02 & 0.1288E-02 \\
 20. & -1.005 &  0.75 & 0.1708E-02 & 0.1430E-02 \\
 20. & -0.755 &  1.00 & 0.1715E-02 & 0.1417E-02 \\
 20. & -0.505 &  1.25 & 0.1252E-02 & 0.9296E-03 \\
 \hline
 30. & -3.005 & -1.25 & 0.1659E-08 & 0.3110E-09 \\
 30. & -2.755 & -1.00 & 0.4690E-05 & 0.1743E-05 \\
 30. & -2.505 & -0.75 & 0.4996E-04 & 0.2469E-04 \\
 30. & -2.255 & -0.50 & 0.1766E-03 & 0.1026E-03 \\
 30. & -2.005 & -0.25 & 0.3887E-03 & 0.2491E-03 \\
 30. & -1.755 &  0.00 & 0.6654E-03 & 0.4509E-03 \\
 30. & -1.505 &  0.25 & 0.9619E-03 & 0.6734E-03 \\
 30. & -1.255 &  0.50 & 0.1222E-02 & 0.8779E-03 \\
 30. & -1.005 &  0.75 & 0.1401E-02 & 0.1023E-02 \\
 30. & -0.755 &  1.00 & 0.1409E-02 & 0.1022E-02 \\
 30. & -0.505 &  1.25 & 0.5914E-03 & 0.3246E-03 \\
 \hline
\end{tabular}
  \end{center}
\end{table}

\newpage
\thispagestyle{empty}
\begin{table}[H]
  \begin{center}
\begin{tabular}{|c|c|c||c|c|}
\hline
$p_{t,W}$ (GeV) & $y_{W,lab}$  & $y_{W,cms}$  & DIS: $W^+$ production & DIS:
$W^-$ production \\
\hline \hline
 40. & -3.005 & -1.25 & --- & --- \\
 40. & -2.755 & -1.00 & 0.1124E-05 & 0.3616E-06 \\
 40. & -2.505 & -0.75 & 0.2090E-04 & 0.9162E-05 \\
 40. & -2.255 & -0.50 & 0.9266E-04 & 0.4768E-04 \\
 40. & -2.005 & -0.25 & 0.2343E-03 & 0.1325E-03 \\
 40. & -1.755 &  0.00 & 0.4375E-03 & 0.2632E-03 \\
 40. & -1.505 &  0.25 & 0.6741E-03 & 0.4212E-03 \\
\hline
\end{tabular}
  \end{center}
\end{table}

\newpage
\thispagestyle{empty}
\begin{table}[H]
  \begin{center}
\begin{tabular}{|c|c|c||c|c|}
\hline
$p_{t,W}$ (GeV) & $y_{W,lab}$  & $y_{W,cms}$  & DIS: $W^+$ production & DIS:
$W^-$ production \\
\hline \hline
 80. & -3.005 & -1.25 & --- & --- \\
 80. & -2.755 & -1.00 & --- & --- \\
 80. & -2.505 & -0.75 & 0.7232E-07 & 0.1910E-07 \\
 80. & -2.255 & -0.50 & 0.2632E-05 & 0.9281E-06 \\
 80. & -2.005 & -0.25 & 0.1510E-04 & 0.6154E-05 \\
 80. & -1.755 &  0.00 & 0.4303E-04 & 0.1909E-04 \\
 80. & -1.505 &  0.25 & 0.8141E-04 & 0.3819E-04 \\
 80. & -1.255 &  0.50 & 0.1077E-03 & 0.5168E-04 \\
 80. & -1.005 &  0.75 & 0.7082E-04 & 0.3033E-04 \\
 80. & -0.755 &  1.00 & --- & --- \\
 80. & -0.505 &  1.25 & --- & --- \\
 \hline
 90. & -3.005 & -1.25 & --- & --- \\
 90. & -2.755 & -1.00 & --- & --- \\
 90. & -2.505 & -0.75 & 0.3538E-08 & 0.7691E-09 \\
 90. & -2.255 & -0.50 & 0.6903E-06 & 0.2202E-06 \\
 90. & -2.005 & -0.25 & 0.5895E-05 & 0.2236E-05 \\
 90. & -1.755 &  0.00 & 0.1951E-04 & 0.8173E-05 \\
 90. & -1.505 &  0.25 & 0.3826E-04 & 0.1690E-04 \\
 90. & -1.255 &  0.50 & 0.4519E-04 & 0.1981E-04 \\
 90. & -1.005 &  0.75 & 0.1100E-04 & 0.3613E-05 \\
 90. & -0.755 &  1.00 & --- & --- \\
 90. & -0.505 &  1.25 & --- & --- \\
 \hline
100. & -3.005 & -1.25 & --- & --- \\
100. & -2.755 & -1.00 & --- & --- \\
100. & -2.505 & -0.75 & 0.1656E-11 & 0.1663E-12 \\
100. & -2.255 & -0.50 & 0.1213E-06 & 0.3464E-07 \\
100. & -2.005 & -0.25 & 0.1918E-05 & 0.6756E-06 \\
100. & -1.755 &  0.00 & 0.7739E-05 & 0.3034E-05 \\
100. & -1.505 &  0.25 & 0.1536E-04 & 0.6341E-05 \\
100. & -1.255 &  0.50 & 0.1384E-04 & 0.5356E-05 \\
100. & -1.005 &  0.75 & 0.2668E-07 & 0.4547E-08 \\
100. & -0.755 &  1.00 & --- & --- \\
100. & -0.505 &  1.25 & --- & --- \\
 \hline
\hline                            
\end{tabular}
\parbox{0.9\textwidth}{\caption{\it
{${d^2\sigma}/{d p_{t,W} \, d y_W}$ (in $pb$/GeV) in LO.}}}
  \end{center}
\end{table}

\newpage
\subsection{Double Differential Cross Sections including NLO}

For the proton CTEQ4M densities~\cite{cteq} are used with NLO strong
coupling ($\Lambda^{(\MS)}_5=202$~MeV).  The NLO direct part is the sum
of 1-jet and 2-jet configurations and thus independent of the jet
definition. The following tables also contain the LO cross sections for
all processes, now using CTEQ4M densities with NLO $\alpha_s$ as for the
NLO direct part. The total sum is given as the sum of the resolved LO,
direct NLO and DIS LO part.  The following values should be used for the
reweighting according to Eqs. (\ref{weight},\ref{weightEPVEC}).  Note
that in order to avoid double counting the cross  sections for LO
resolved photoproduction convoluted with NLO parton densities have to be
added to the NLO direct part.

\subsubsection{$W^+$ production in $e^+p$ scattering including NLO at $E_p=920$~GeV}

\begin{table}[H]
\hspace{-1.2cm}
\begin{tabular}{|c|c|c||c|c|c|c||c|}
\hline
$p_{t,W}$ (GeV) & $y_{W,lab}$  & $y_{W,cms}$ & resolved LO & direct LO &
direct NLO & DIS LO & total \\
\hline \hline
 10. & -3.005 & -1.25 & 0.5250E-06 & 0.1196E-05 & 0.1920E-05 & 0.3387E-06 & 0.2784E-05 \\
 10. & -2.755 & -1.00 & 0.8354E-04 & 0.1356E-03 & 0.1740E-03 & 0.3761E-04 & 0.2951E-03 \\
 10. & -2.505 & -0.75 & 0.5007E-03 & 0.7714E-03 & 0.8789E-03 & 0.2158E-03 & 0.1595E-02 \\
 10. & -2.255 & -0.50 & 0.1233E-02 & 0.1938E-02 & 0.2043E-02 & 0.5535E-03 & 0.3830E-02 \\
 10. & -2.005 & -0.25 & 0.2032E-02 & 0.3359E-02 & 0.3362E-02 & 0.9887E-03 & 0.6383E-02 \\
 10. & -1.755 &  0.00 & 0.2666E-02 & 0.4720E-02 & 0.4537E-02 & 0.1439E-02 & 0.8642E-02 \\
 10. & -1.505 &  0.25 & 0.2990E-02 & 0.5780E-02 & 0.5425E-02 & 0.1833E-02 & 0.1025E-01 \\
\hline                            
\end{tabular}
\end{table}

\newpage
\thispagestyle{empty}
\begin{table}[H]
\hspace{-1.2cm}
\begin{tabular}{|c|c|c||c|c|c|c||c|}
\hline
$p_{t,W}$ (GeV) & $y_{W,lab}$  & $y_{W,cms}$ & resolved LO & direct LO &
direct NLO & DIS LO & total \\
\hline \hline
 30. & -3.005 & -1.25 & 0.1256E-09 & 0.1821E-08 & 0.2969E-08 & 0.8648E-09 & 0.3959E-08 \\
 30. & -2.755 & -1.00 & 0.1156E-05 & 0.8038E-05 & 0.1097E-04 & 0.3385E-05 & 0.1551E-04 \\
 30. & -2.505 & -0.75 & 0.1467E-04 & 0.9825E-04 & 0.1253E-03 & 0.4077E-04 & 0.1807E-03 \\
 30. & -2.255 & -0.50 & 0.5033E-04 & 0.3745E-03 & 0.4580E-03 & 0.1552E-03 & 0.6636E-03 \\
 30. & -2.005 & -0.25 & 0.9932E-04 & 0.8685E-03 & 0.1031E-02 & 0.3646E-03 & 0.1495E-02 \\
 30. & -1.755 &  0.00 & 0.1442E-03 & 0.1529E-02 & 0.1776E-02 & 0.6563E-03 & 0.2576E-02 \\
 30. & -1.505 &  0.25 & 0.1693E-03 & 0.2252E-02 & 0.2566E-02 & 0.9994E-03 & 0.3735E-02 \\
\hline                            
\end{tabular}
\end{table}

\newpage
\thispagestyle{empty}
\begin{table}[H]
\hspace{-1.2cm}
\begin{tabular}{|c|c|c||c|c|c|c||c|}
\hline
$p_{t,W}$ (GeV) & $y_{W,lab}$  & $y_{W,cms}$ & resolved LO & direct LO &
direct NLO & DIS LO & total \\
\hline \hline
 70. & -3.005 & -1.25 &   ---      &   ---      &   ---      &   ---      &   ---      \\
 70. & -2.755 & -1.00 & 0.1961E-12 & 0.2683E-10 & 0.3963E-10 & 0.1850E-10 & 0.5833E-10 \\
 70. & -2.505 & -0.75 & 0.1529E-07 & 0.5882E-06 & 0.6953E-06 & 0.3481E-06 & 0.1059E-05 \\
 70. & -2.255 & -0.50 & 0.2514E-06 & 0.1046E-04 & 0.1183E-04 & 0.5930E-05 & 0.1801E-04 \\
 70. & -2.005 & -0.25 & 0.9348E-06 & 0.5033E-04 & 0.5559E-04 & 0.2797E-04 & 0.8450E-04 \\
 70. & -1.755 &  0.00 & 0.1787E-05 & 0.1371E-03 & 0.1489E-03 & 0.7516E-04 & 0.2259E-03 \\
 70. & -1.505 &  0.25 & 0.2184E-05 & 0.2662E-03 & 0.2859E-03 & 0.1460E-03 & 0.4341E-03 \\
\hline                            
\hline                            
\end{tabular}
\parbox{0.9\textwidth}{\caption{\it
{${d^2\sigma}/{d p_{t,W} \, d y_W}$ (in $pb$/GeV) in LO and NLO.}}}
\end{table}

\subsubsection{$W^-$ production in $e^+p$ scattering including NLO at $E_p=920$~GeV}

\begin{table}[H]
\hspace{-1.2cm}
\begin{tabular}{|c|c|c||c|c|c|c||c|}
\hline
$p_{t,W}$ (GeV) & $y_{W,lab}$  & $y_{W,cms}$ & resolved LO & direct LO &
direct NLO & DIS LO & total \\
\hline \hline
 10. & -3.005 & -1.25 & 0.1134E-06 & 0.3140E-06 & 0.5113E-06 & 0.8955E-07 & 0.7143E-06 \\
 10. & -2.755 & -1.00 & 0.3554E-04 & 0.6458E-04 & 0.8387E-04 & 0.1809E-04 & 0.1375E-03 \\
 10. & -2.505 & -0.75 & 0.3109E-03 & 0.4974E-03 & 0.5700E-03 & 0.1402E-03 & 0.1021E-02 \\
 10. & -2.255 & -0.50 & 0.9940E-03 & 0.1518E-02 & 0.1595E-02 & 0.4328E-03 & 0.3022E-02 \\
 10. & -2.005 & -0.25 & 0.1968E-02 & 0.2992E-02 & 0.2949E-02 & 0.8679E-03 & 0.5785E-02 \\
 10. & -1.755 &  0.00 & 0.2922E-02 & 0.4552E-02 & 0.4258E-02 & 0.1337E-02 & 0.8517E-02 \\
 10. & -1.505 &  0.25 & 0.3541E-02 & 0.5820E-02 & 0.5285E-02 & 0.1744E-02 & 0.1057E-01 \\
\hline                            
\end{tabular}
\end{table}

\newpage
\thispagestyle{empty}
\begin{table}[H]
\hspace{-1.2cm}
\begin{tabular}{|c|c|c||c|c|c|c||c|}
\hline
$p_{t,W}$ (GeV) & $y_{W,lab}$  & $y_{W,cms}$ & resolved LO & direct LO &
direct NLO & DIS LO & total \\
\hline \hline
 40. & -3.005 & -1.25 &   ---      &   ---      &   ---      &   ---      &   ---      \\
 40. & -2.755 & -1.00 & 0.4289E-07 & 0.5102E-06 & 0.6913E-06 & 0.2409E-06 & 0.9751E-06 \\
 40. & -2.505 & -0.75 & 0.1559E-05 & 0.1531E-04 & 0.1940E-04 & 0.6896E-05 & 0.2786E-04 \\
 40. & -2.255 & -0.50 & 0.8955E-05 & 0.8660E-04 & 0.1058E-03 & 0.3828E-04 & 0.1530E-03 \\
 40. & -2.005 & -0.25 & 0.2434E-04 & 0.2525E-03 & 0.3000E-03 & 0.1109E-03 & 0.4353E-03 \\
 40. & -1.755 &  0.00 & 0.4337E-04 & 0.5152E-03 & 0.6011E-03 & 0.2251E-03 & 0.8696E-03 \\
 40. & -1.505 &  0.25 & 0.5720E-04 & 0.8358E-03 & 0.9605E-03 & 0.3675E-03 & 0.1385E-02 \\
\hline                            
\end{tabular}
\end{table}

\begin{table}[H]
\hspace{-1.2cm}
\begin{tabular}{|c|c|c||c|c|c|c||c|}
\hline
$p_{t,W}$ (GeV) & $y_{W,lab}$  & $y_{W,cms}$ & resolved LO & direct LO &
direct NLO & DIS LO & total \\
\hline \hline
 80. & -3.005 & -1.25 &   ---      &   ---      &   ---      &   ---      &   ---      \\
 80. & -2.755 & -1.00 &   ---      &   ---      &   ---      &   ---      &   ---      \\
 80. & -2.505 & -0.75 & 0.3404E-09 & 0.1686E-07 & 0.2042E-07 & 0.9790E-08 & 0.3055E-07 \\
 80. & -2.255 & -0.50 & 0.2422E-07 & 0.1013E-05 & 0.1161E-05 & 0.5515E-06 & 0.1737E-05 \\
 80. & -2.005 & -0.25 & 0.1596E-06 & 0.7422E-05 & 0.8298E-05 & 0.3893E-05 & 0.1235E-04 \\
 80. & -1.755 &  0.00 & 0.4057E-06 & 0.2433E-04 & 0.2686E-04 & 0.1254E-04 & 0.3981E-04 \\
 80. & -1.505 &  0.25 & 0.5471E-06 & 0.4989E-04 & 0.5459E-04 & 0.2540E-04 & 0.8053E-04 \\
\hline                            
\hline                            
\end{tabular}
\parbox{0.9\textwidth}{\caption{\it
{${d^2\sigma}/{d p_{t,W} \, d y_W}$ (in $pb$/GeV) in LO and NLO.}}}
\end{table}

\subsubsection{$W^+$ production in $e^+p$ scattering including NLO at $E_p=820$~GeV}

\begin{table}[H]
\hspace{-1.2cm}
\begin{tabular}{|c|c|c||c|c|c|c||c|}
\hline
$p_{t,W}$ (GeV) & $y_{W,lab}$  & $y_{W,cms}$ & resolved LO & direct LO &
direct NLO & DIS LO & total \\
\hline \hline
 10. & -2.948 & -1.25 & 0.1812E-07 & 0.5665E-07 & 0.1010E-06 & 0.1650E-07 & 0.1356E-06 \\
 10. & -2.698 & -1.00 & 0.3865E-04 & 0.6791E-04 & 0.9051E-04 & 0.1899E-04 & 0.1482E-03 \\
 10. & -2.448 & -0.75 & 0.3362E-03 & 0.5431E-03 & 0.6354E-03 & 0.1527E-03 & 0.1124E-02 \\
 10. & -2.198 & -0.50 & 0.9483E-03 & 0.1545E-02 & 0.1662E-02 & 0.4429E-03 & 0.3053E-02 \\
 10. & -1.948 & -0.25 & 0.1664E-02 & 0.2840E-02 & 0.2890E-02 & 0.8382E-03 & 0.5392E-02 \\
 10. & -1.698 &  0.00 & 0.2251E-02 & 0.4116E-02 & 0.4020E-02 & 0.1258E-02 & 0.7529E-02 \\
 10. & -1.448 &  0.25 & 0.2557E-02 & 0.5124E-02 & 0.4887E-02 & 0.1630E-02 & 0.9074E-02 \\
\hline                            
\end{tabular}
\end{table}

\newpage
\thispagestyle{empty}
\begin{table}[H]
\hspace{-1.2cm}
\begin{tabular}{|c|c|c||c|c|c|c||c|}
\hline
$p_{t,W}$ (GeV) & $y_{W,lab}$  & $y_{W,cms}$ & resolved LO & direct LO &
direct NLO & DIS LO & total \\
\hline \hline
 40. & -2.948 & -1.25 &   ---      &   ---      &   ---      &   ---      &   ---      \\
 40. & -2.698 & -1.00 & 0.2833E-07 & 0.3964E-06 & 0.5436E-06 & 0.1924E-06 & 0.7644E-06 \\
 40. & -2.448 & -0.75 & 0.1613E-05 & 0.1927E-04 & 0.2430E-04 & 0.8926E-05 & 0.3484E-04 \\
 40. & -2.198 & -0.50 & 0.8909E-05 & 0.1177E-03 & 0.1422E-03 & 0.5431E-04 & 0.2054E-03 \\
 40. & -1.948 & -0.25 & 0.2184E-04 & 0.3501E-03 & 0.4107E-03 & 0.1622E-03 & 0.5947E-03 \\
 40. & -1.698 &  0.00 & 0.3519E-04 & 0.7213E-03 & 0.8279E-03 & 0.3376E-03 & 0.1201E-02 \\
 40. & -1.448 &  0.25 & 0.4259E-04 & 0.1182E-02 & 0.1332E-02 & 0.5657E-03 & 0.1940E-02 \\
\hline                            
\end{tabular}
\end{table}

\begin{table}[H]
\hspace{-1.2cm}
\begin{tabular}{|c|c|c||c|c|c|c||c|}
\hline
$p_{t,W}$ (GeV) & $y_{W,lab}$  & $y_{W,cms}$ & resolved LO & direct LO &
direct NLO & DIS LO & total \\
\hline \hline
 80. & -2.948 & -1.25 &   ---      &   ---      &   ---      &   ---      &   ---      \\
 80. & -2.698 & -1.00 &   ---      &   ---      &   ---      &   ---      &   ---      \\
 80. & -2.448 & -0.75 & 0.9523E-10 & 0.8224E-08 & 0.1019E-07 & 0.5227E-08 & 0.1551E-07 \\
 80. & -2.198 & -0.50 & 0.1875E-07 & 0.1331E-05 & 0.1515E-05 & 0.7907E-06 & 0.2324E-05 \\
 80. & -1.948 & -0.25 & 0.1315E-06 & 0.1186E-04 & 0.1308E-04 & 0.6804E-05 & 0.2001E-04 \\
 80. & -1.698 &  0.00 & 0.3096E-06 & 0.4212E-04 & 0.4567E-04 & 0.2376E-04 & 0.6974E-04 \\
 80. & -1.448 &  0.25 & 0.3669E-06 & 0.9032E-04 & 0.9697E-04 & 0.5041E-04 & 0.1478E-03 \\
\hline                            
\hline                            
\end{tabular}
\parbox{0.9\textwidth}{\caption{\it
{${d^2\sigma}/{d p_{t,W} \, d y_W}$ (in $pb$/GeV) in LO and NLO.}}}
\end{table}

\subsubsection{$W^-$ production in $e^+p$ scattering including NLO at $E_p=820$~GeV}

\begin{table}[H]
\hspace{-1.2cm}
\begin{tabular}{|c|c|c||c|c|c|c||c|}
\hline
$p_{t,W}$ (GeV) & $y_{W,lab}$  & $y_{W,cms}$ & resolved LO & direct LO &
direct NLO & DIS LO & total \\
\hline \hline
 10. & -2.948 & -1.25 & 0.3004E-08 & 0.1140E-07 & 0.2057E-07 & 0.3341E-08 & 0.2692E-07 \\
 10. & -2.698 & -1.00 & 0.1487E-04 & 0.2928E-04 & 0.3954E-04 & 0.8278E-05 & 0.6269E-04 \\
 10. & -2.448 & -0.75 & 0.1964E-03 & 0.3295E-03 & 0.3882E-03 & 0.9335E-04 & 0.6779E-03 \\
 10. & -2.198 & -0.50 & 0.7325E-03 & 0.1161E-02 & 0.1244E-02 & 0.3320E-03 & 0.2308E-02 \\
 10. & -1.948 & -0.25 & 0.1565E-02 & 0.2456E-02 & 0.2467E-02 & 0.7127E-03 & 0.4745E-02 \\
 10. & -1.698 &  0.00 & 0.2420E-02 & 0.3888E-02 & 0.3708E-02 & 0.1144E-02 & 0.7272E-02 \\
 10. & -1.448 &  0.25 & 0.2991E-02 & 0.5083E-02 & 0.4701E-02 & 0.1523E-02 & 0.9215E-02 \\
\hline                            
\end{tabular}
\end{table}

\newpage
\thispagestyle{empty}
\begin{table}[H]
\hspace{-1.2cm}
\begin{tabular}{|c|c|c||c|c|c|c||c|}
\hline
$p_{t,W}$ (GeV) & $y_{W,lab}$  & $y_{W,cms}$ & resolved LO & direct LO &
direct NLO & DIS LO & total \\
\hline \hline
 40. & -2.948 & -1.25 &   ---      &   ---      &   ---      &   ---      &   ---      \\
 40. & -2.698 & -1.00 & 0.7898E-08 & 0.1112E-06 & 0.1550E-06 & 0.5360E-07 & 0.2165E-06 \\
 40. & -2.448 & -0.75 & 0.7347E-06 & 0.7832E-05 & 0.1007E-04 & 0.3554E-05 & 0.1435E-04 \\
 40. & -2.198 & -0.50 & 0.5540E-05 & 0.5679E-04 & 0.6994E-04 & 0.2526E-04 & 0.1007E-03 \\
 40. & -1.948 & -0.25 & 0.1696E-04 & 0.1853E-03 & 0.2221E-03 & 0.8175E-04 & 0.3208E-03 \\
 40. & -1.698 &  0.00 & 0.3198E-04 & 0.4013E-03 & 0.4716E-03 & 0.1760E-03 & 0.6796E-03 \\
 40. & -1.448 &  0.25 & 0.4296E-04 & 0.6725E-03 & 0.7771E-03 & 0.2952E-03 & 0.1115E-02 \\
\hline                            
\end{tabular}
\end{table}

\begin{table}[H]
\hspace{-1.2cm}
\begin{tabular}{|c|c|c||c|c|c|c||c|}
\hline
$p_{t,W}$ (GeV) & $y_{W,lab}$  & $y_{W,cms}$ & resolved LO & direct LO &
direct NLO & DIS LO & total \\
\hline \hline
 80. & -2.948 & -1.25 &   ---      &   ---      &   ---      &   ---      &   ---      \\
 80. & -2.698 & -1.00 &   ---      &   ---      &   ---      &   ---      &   ---      \\
 80. & -2.448 & -0.75 & 0.2296E-10 & 0.1616E-08 & 0.2051E-08 & 0.9571E-09 & 0.3031E-08 \\
 80. & -2.198 & -0.50 & 0.7756E-08 & 0.3845E-06 & 0.4486E-06 & 0.2089E-06 & 0.6653E-06 \\
 80. & -1.948 & -0.25 & 0.7353E-07 & 0.3950E-05 & 0.4478E-05 & 0.2077E-05 & 0.6629E-05 \\
 80. & -1.698 &  0.00 & 0.2101E-06 & 0.1474E-04 & 0.1645E-04 & 0.7569E-05 & 0.2423E-04 \\
 80. & -1.448 &  0.25 & 0.2790E-06 & 0.3117E-04 & 0.3451E-04 & 0.1583E-04 & 0.5062E-04 \\
\hline                            
\hline                            
\end{tabular}
\parbox{0.9\textwidth}{\caption{\it
{${d^2\sigma}/{d p_{t,W} \, d y_W}$ (in $pb$/GeV) in LO and NLO.}}}
\end{table}

\newpage

\subsubsection{$W^\pm$ production in $e^-p$ scattering at $E_p=920$~GeV}

In $e^-p$ scattering the results for resolved and direct photoproduction
are identical to those of $e^+p$ scattering, while the DIS results are
different:

\begin{table}[H]
  \begin{center}
\begin{tabular}{|c|c|c||c|c|}
\hline
$p_{t,W}$ (GeV) & $y_{W,lab}$  & $y_{W,cms}$  & DIS LO: $W^+$ production &
DIS LO:
$W^-$ production \\
\hline \hline
 10. & -3.005 & -1.25 & 0.3420E-06 & 0.9027E-07 \\
 10. & -2.755 & -1.00 & 0.3762E-04 & 0.1820E-04 \\
 10. & -2.505 & -0.75 & 0.2142E-03 & 0.1411E-03 \\
 10. & -2.255 & -0.50 & 0.5454E-03 & 0.4350E-03 \\
 10. & -2.005 & -0.25 & 0.9589E-03 & 0.8715E-03 \\
 10. & -1.755 &  0.00 & 0.1383E-02 & 0.1349E-02 \\
 10. & -1.505 &  0.25 & 0.1733E-02 & 0.1756E-02 \\
 10. & -1.255 &  0.50 & 0.1961E-02 & 0.2016E-02 \\
 10. & -1.005 &  0.75 & 0.2047E-02 & 0.2094E-02 \\
 10. & -0.755 &  1.00 & 0.1966E-02 & 0.1956E-02 \\
 10. & -0.505 &  1.25 & 0.1502E-02 & 0.1371E-02 \\
 \hline
 20. & -3.005 & -1.25 & 0.3355E-07 & 0.7616E-08 \\
 20. & -2.755 & -1.00 & 0.1202E-04 & 0.5075E-05 \\
 20. & -2.505 & -0.75 & 0.9451E-04 & 0.5322E-04 \\
 20. & -2.255 & -0.50 & 0.2886E-03 & 0.1929E-03 \\
 20. & -2.005 & -0.25 & 0.5812E-03 & 0.4328E-03 \\
 20. & -1.755 &  0.00 & 0.9301E-03 & 0.7368E-03 \\
 20. & -1.505 &  0.25 & 0.1276E-02 & 0.1043E-02 \\
 20. & -1.255 &  0.50 & 0.1567E-02 & 0.1304E-02 \\
 20. & -1.005 &  0.75 & 0.1755E-02 & 0.1464E-02 \\
 20. & -0.755 &  1.00 & 0.1773E-02 & 0.1464E-02 \\
 20. & -0.505 &  1.25 & 0.1288E-02 & 0.9484E-03 \\
 \hline
 30. & -3.005 & -1.25 & 0.9148E-09 & 0.1647E-09 \\
 30. & -2.755 & -1.00 & 0.3540E-05 & 0.1310E-05 \\
 30. & -2.505 & -0.75 & 0.4188E-04 & 0.2064E-04 \\
 30. & -2.255 & -0.50 & 0.1568E-03 & 0.9077E-04 \\
 30. & -2.005 & -0.25 & 0.3606E-03 & 0.2305E-03 \\
 30. & -1.755 &  0.00 & 0.6351E-03 & 0.4291E-03 \\
 30. & -1.505 &  0.25 & 0.9401E-03 & 0.6563E-03 \\
 30. & -1.255 &  0.50 & 0.1217E-02 & 0.8734E-03 \\
 30. & -1.005 &  0.75 & 0.1419E-02 & 0.1031E-02 \\
 30. & -0.755 &  1.00 & 0.1434E-02 & 0.1036E-02 \\
 30. & -0.505 &  1.25 & 0.5798E-03 & 0.3126E-03 \\
 \hline
\end{tabular}
  \end{center}
\end{table}

\newpage
\thispagestyle{empty}
\begin{table}[H]
  \begin{center}
\begin{tabular}{|c|c|c||c|c|}
\hline
$p_{t,W}$ (GeV) & $y_{W,lab}$  & $y_{W,cms}$  & DIS LO: $W^+$ production & DIS LO:
$W^-$ production \\
\hline \hline
 40. & -3.005 & -1.25 & --- & --- \\
 40. & -2.755 & -1.00 & 0.8133E-06 & 0.2612E-06 \\
 40. & -2.505 & -0.75 & 0.1708E-04 & 0.7477E-05 \\
 40. & -2.255 & -0.50 & 0.8089E-04 & 0.4150E-04 \\
 40. & -2.005 & -0.25 & 0.2134E-03 & 0.1206E-03 \\
 40. & -1.755 &  0.00 & 0.4127E-03 & 0.2463E-03 \\
 40. & -1.505 &  0.25 & 0.6501E-03 & 0.4047E-03 \\
\hline
\end{tabular}
  \end{center}
\end{table}

\newpage
\thispagestyle{empty}
\begin{table}[H]
  \begin{center}
\begin{tabular}{|c|c|c||c|c|}
\hline
$p_{t,W}$ (GeV) & $y_{W,lab}$  & $y_{W,cms}$  & DIS LO: $W^+$ production & DIS LO:
$W^-$ production \\
\hline \hline
 80. & -3.005 & -1.25 & --- & --- \\
 80. & -2.755 & -1.00 & --- & --- \\
 80. & -2.505 & -0.75 & 0.4970E-07 & 0.1302E-07 \\
 80. & -2.255 & -0.50 & 0.2066E-05 & 0.7305E-06 \\
 80. & -2.005 & -0.25 & 0.1273E-04 & 0.5169E-05 \\
 80. & -1.755 &  0.00 & 0.3776E-04 & 0.1672E-04 \\
 80. & -1.505 &  0.25 & 0.7347E-04 & 0.3424E-04 \\
 80. & -1.255 &  0.50 & 0.9855E-04 & 0.4680E-04 \\
 80. & -1.005 &  0.75 & 0.6363E-04 & 0.2688E-04 \\
 80. & -0.755 &  1.00 & --- & --- \\
 80. & -0.505 &  1.25 & --- & --- \\
 \hline
 90. & -3.005 & -1.25 & --- & --- \\
 90. & -2.755 & -1.00 & --- & --- \\
 90. & -2.505 & -0.75 & 0.2200E-08 & 0.4673E-09 \\
 90. & -2.255 & -0.50 & 0.5207E-06 & 0.1661E-06 \\
 90. & -2.005 & -0.25 & 0.4825E-05 & 0.1828E-05 \\
 90. & -1.755 &  0.00 & 0.1674E-04 & 0.6990E-05 \\
 90. & -1.505 &  0.25 & 0.3365E-04 & 0.1479E-04 \\
 90. & -1.255 &  0.50 & 0.4011E-04 & 0.1745E-04 \\
 90. & -1.005 &  0.75 & 0.9159E-05 & 0.2977E-05 \\
 90. & -0.755 &  1.00 & --- & --- \\
 90. & -0.505 &  1.25 & --- & --- \\
 \hline
100. & -3.005 & -1.25 & --- & --- \\
100. & -2.755 & -1.00 & --- & --- \\
100. & -2.505 & -0.75 & 0.6625E-12 & 0.5935E-13 \\
100. & -2.255 & -0.50 & 0.8649E-07 & 0.2458E-07 \\
100. & -2.005 & -0.25 & 0.1516E-05 & 0.5324E-06 \\
100. & -1.755 &  0.00 & 0.6432E-05 & 0.2518E-05 \\
100. & -1.505 &  0.25 & 0.1313E-04 & 0.5396E-05 \\
100. & -1.255 &  0.50 & 0.1178E-04 & 0.4517E-05 \\
100. & -1.005 &  0.75 & 0.1757E-07 & 0.2937E-08 \\
100. & -0.755 &  1.00 & --- & --- \\
100. & -0.505 &  1.25 & --- & --- \\
 \hline
\hline                            
\end{tabular}
\parbox{0.9\textwidth}{\caption{\it
{${d^2\sigma}/{d p_{t,W} \, d y_W}$ (in $pb$/GeV) in LO.}}}
  \end{center}
\end{table}

\newpage
\section{Cross Section for $p_{t,W} < 5\,{\rm GeV}$}

The cross sections for $p_{t,W} < 5\,{\rm GeV}$ are presented for the
individual $W$ production processes as calculated in Eqs.
(\ref{reslt5gev},\ref{dirlt5gev}) and $\sigma_{res}^{LO}(p_{t,W} <
5\,{\rm GeV}) = \sigma_{res}^{LO}({\rm total})$\footnote{See left
diagram in Fig.~\ref{fig:wtot_LOgraphs}.}. We give the LO and NLO
results for proton beam energies of $E_p=820$~GeV and $E_p=920$~GeV (and
electron/positron beam energies of $E_e = 27.5$~GeV). The
ACFGP~\cite{acfgp} parton densities are chosen for the photon. All other
settings and numerical values of parameters are as in~\cite{DSS}. 

\subsection{Cross Section for $p_{t,W} < 5\,{\rm GeV}$ in LO}

For the proton CTEQ4L densities~\cite{cteq} are used with LO strong
coupling ($\Lambda^{(LO)}_5=181$~MeV).  The following values can be used
for consistency checks in LO.

\subsubsection{$e^+p$ scattering}

\begin{table}[H]
  \begin{center}
\begin{tabular}{|c|c||c|c|c||c|}
\hline
charge of $W$ & $E_p$ (GeV) & resolved & direct & DIS & total \\
\hline \hline
 +  & 920 & 0.2286 & -0.1669 & 0.0178 & 0.0794 \\
\hline                            
 -- & 920 & 0.2746 & -0.2080 & 0.0196 & 0.0861 \\
\hline                            
 +  & 820 & 0.1964 & -0.1433 & 0.0153 & 0.0683 \\
\hline                            
 -- & 820 & 0.2344 & -0.1776 & 0.0167 & 0.0735 \\
\hline                            
\hline                            
\end{tabular}
\parbox{0.9\textwidth}{\caption{\it
{$\sigma(p_{t,W} < 5\,{\rm GeV})$ (in $pb$) in LO.}}}
  \end{center}
\end{table}

\subsubsection{$e^-p$ scattering}

In $e^-p$ scattering the results for resolved and direct photoproduction
are identical to those of $e^+p$ scattering, while the DIS results are
different:

\begin{table}[H]
  \begin{center}
\begin{tabular}{|c|c||c|}
\hline
charge of $W$ & $E_p$ (GeV) & DIS \\
\hline \hline
 +  & 920 & 0.0171 \\
\hline                            
 -- & 920 & 0.0196 \\
\hline                            
\hline                            
\end{tabular}
\parbox{0.9\textwidth}{\caption{\it
{$\sigma(p_{t,W} < 5\,{\rm GeV})$ (in $pb$) in LO.}}}
  \end{center}
\end{table}

\subsection{Cross Section for $p_{t,W} < 5\,{\rm GeV}$ including NLO
corrections}

For the proton CTEQ4M densities~\cite{cteq} are used with NLO strong
coupling ($\Lambda^{(\MS)}_5=202$~MeV).  The latter is also used for the
following LO cross sections.  The total sum is given as the sum of the
resolved NLO, direct LO and DIS LO part.  The following values should be
used for the reweighting according to Eqs.
(\ref{weightlbin},\ref{weightlbinEPVEC}). 

\subsubsection{$e^+p$ scattering}

\begin{table}[H]
  \begin{center}
\begin{tabular}{|c|c||c|c|c|c||c|}
\hline
charge of $W$ & $E_p$ (GeV) & resolved LO & resolved NLO & direct LO & DIS
LO & total \\
\hline
\hline                            
 +  & 920 & 0.2326 & 0.3435 & -0.1708 & 0.0181 & 0.1908 \\
\hline                            
 -- & 920 & 0.2804 & 0.4173 & -0.2133 & 0.0200 & 0.2240 \\
\hline                            
 +  & 820 & 0.1990 & 0.2952 & -0.1459 & 0.0156 & 0.1649 \\
\hline                            
 -- & 820 & 0.2384 & 0.3567 & -0.1813 & 0.0170 & 0.1924 \\
\hline                            
\hline                            
\end{tabular}
\parbox{0.9\textwidth}{\caption{\it
{$\sigma(p_{t,W} < 5\,{\rm GeV})$ (in $pb$) in LO and NLO.}}}
  \end{center}
\end{table}

\subsubsection{$e^-p$ scattering}

In $e^-p$ scattering the results for resolved and direct photoproduction
are identical to those of $e^+p$ scattering, while the DIS results are
different:

\begin{table}[H]
  \begin{center}
\begin{tabular}{|c|c||c|}
\hline
charge of $W$ & $E_p$ (GeV) & DIS LO \\
\hline
\hline                            
 +  & 920 & 0.0174 \\
\hline                            
 -- & 920 & 0.0201 \\
\hline                            
\hline                            
\end{tabular}
\parbox{0.9\textwidth}{\caption{\it
{$\sigma(p_{t,W} < 5\,{\rm GeV})$ (in $pb$) in LO and NLO.}}}
  \end{center}
\end{table}

\end{appendix}

\newpage


\end{document}